\documentclass[article,twocolumn,preprintnumbers,nofootinbib]{revtex4-1}
\usepackage{enumerate}
\usepackage{mathrsfs,amsmath,amssymb}
\usepackage{natbib}
\usepackage{graphicx}
\usepackage{cancel}
\usepackage[normalem]{ulem}
\usepackage[pdftex,bookmarks,linktocpage,pdfpagelabels,plainpages=false,hyperfigures,linkcolor=blue,citecolor=blue]{hyperref} 
\hypersetup{colorlinks=true}

\newcommand\be{\begin{equation}}
\newcommand\ee{\end{equation}}
\newcommand\bea{\begin{eqnarray}}
\newcommand\eea{\end{eqnarray}}

\begin{document}
\bibliographystyle{apsrev4-1}


\title{First Order Color  Symmetry Breaking and Restoration Triggered by Electroweak Symmetry Non-restoration}
\author{Wei Chao$^1$}
\email{chaowei@bnu.edu.cn}
\author{Huai-Ke Guo$^2$}
\email{huaike.guo@utah.edu}
\author{Xiu-Fei Li$^1$}
\email{xiufeili@mail.bnu.edu.cn}
\affiliation{$^1$Center for Advanced Quantum Studies, Department of Physics, Beijing Normal University, Beijing, 100875, China \\
$^2$ Department of Physics and Astronomy, University of Utah, Salt Lake City, UT 84112, USA}

\begin{abstract}
In this paper we propose a new approach for the spontaneous breaking and restoration of the $SU(3)_C$ color symmetry in the framework of  electroweak symmetry non-restoration (EWSNR) at high temperature, which provides an alternative approach for the Baryogenesis. Due to the exotic high vacuum expectation value (VEV) of the SM Higgs doublet  in EWSNR, the color symmetry can be spontaneous broken succeeding the electroweak phase transition  whenever there is a negative quartic coupling between the SM Higgs and a scalar color triplet.  The color symmetry is then restored at low temperature as the VEV of SM Higgs evolving to small value. We show that the phase transitions related to color breaking and restoration can be first order, and the stochastic gravitational wave (GW) signals are smoking-gun of these processes.  We demonstrate the possibility of detecting these GW signals in future GW experiments such as DECIGO and BBO.
\end{abstract}

\maketitle

\section{Introduction}\label{1}
The discovery of the  Standard Model (SM) Higgs~\cite{ATLAS:2012yve,CMS:2012qbp} and measurements of Higgs couplings with the second~\cite{CMS:2020xwi,ATLAS:2020fzp} and third generation~\cite{ATLAS:2015xst,CMS:2017zyp} fermions at the CERN LHC are milestones of the high energy physics, which confirms the correctness of Higgs mechanism and leaves examinations of the Higgs property as well as the pattern of the electroweak symmetry breaking urgent tasks. It is well-known that the SM of particle physics is not a final theory since it can not explain the matter-antimatter asymmetry of the Universe (BAU)~\cite{Dine:2003ax} as well as the cold dark matter~\cite{Jungman:1995df,Bertone:2004pz}.  Higgs is an important portal to new physics beyond the SM.

Of various extensions to the SM Higgs sector,  the scenario with strongly first order phase transition is attractive since it may both generate the BAU via the electroweak Baryogenesis mechanism (EWBG)~\cite{Morrissey:2012db} and gives rise to a stochastic gravitational wave (GW) signal which can be tested by future space based interferometer~\cite{Caprini:2015zlo}.  In addition, dark matter can be addressed in Higgs portal models~\cite{Patt:2006fw}.  Even though EWBG is a promising mechanism for BAU, it suffers from tension between the requirement of a large CP violation as required by the Sakharov condition~\cite{Sakharov:1967dj} and the non-observation of electron or nucleon electric dipole moments (EDM)~\cite{ACME:2018yjb,Chupp:2017rkp} in low energy precision measurement experiments.  One possible way out is lifting up the scale of electroweak symmetry breaking such that the sphaleron is quenched at high energy scale, which results in much heavier particles for BAU,  and thus EDM bounds can be avoided.  This can be realized in symmetry non-restoration (SNR) models~\cite{Weinberg:1974hy,Mohapatra:1979qt,Fujimoto:1984hr},  in which additional scalar singlets or fermions interacting with SM Higgs are introduced to provide negative thermal mass for it at very high temperature. In SNR, the vacuum expectation value (VEV) of the SM Higgs doublet can evolve to a value much larger than $246$ GeV, and further up to a cut-off scale $\Lambda$, above which additional Higgs interactions may restore  the electroweak symmetry.   
Actually, investigation of SNR has a long history~\cite{Weinberg:1974hy,Mohapatra:1979qt,Fujimoto:1984hr,Dvali:1995cj,Salomonson:1984rh,Bimonte:1995sc,Dvali:1996zr,Orloff:1996yn,Gavela:1998ux,Ahriche:2010kh,Espinosa:2004pn,Bajc:1998jr,Agrawal:2021alq} and recent studies~\cite{Meade:2018saz,Baldes:2018nel,Glioti:2018roy,Matsedonskyi:2020mlz,Matsedonskyi:2020kuy,Carena:2021onl,Biekotter:2021ysx,Bai:2021hfb,Matsedonskyi:2021hti} show that this scenario contains rich phenomenologies deserving dedicated studies. 

\begin{figure*}[t]
	\includegraphics[width=7cm]{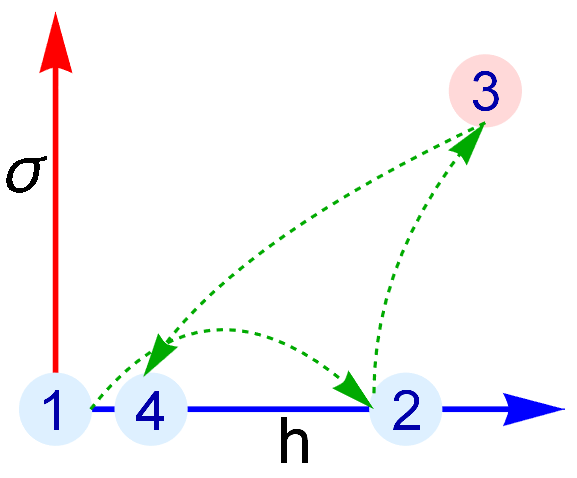}
	\includegraphics[width=8.3cm]{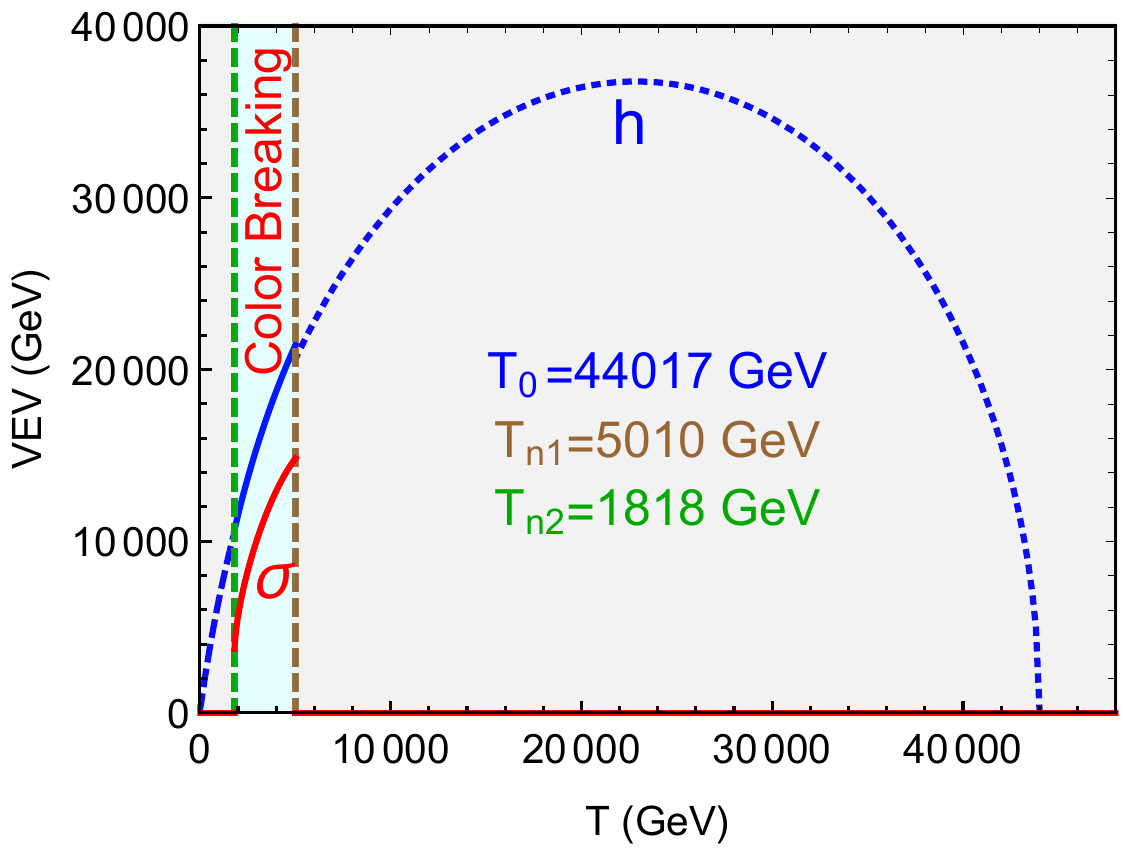}
	\caption{Left-panel: Schematic diagram for phase transitions. Right-panel:  VEVs of the SM Higgs (blue lines) and  color triplet scalar (red line) as the function of temperature for the benchmark point  in TABLE \ref{table1}. Vertical dashed brown(green) line  corresponds to the border of color symmetry breaking(restoration).  $T_{n1}$ and  $T_{n2}$ are the corresponding bubble nucleation temperatures. $T_0$ is the EWPT temperature.}
	\label{vev-hc}
\end{figure*}

In this paper, we investigate the spontaneous breaking of the $SU(3)_C$ gauge symmetry at high temperature. Weinberg~\cite{Weinberg:1974hy} points out that gauge and global symmetries can be broken at high temperature, then restored at low temperature.  How to realize the breaking and restoration of $SU(3)_C$ gauge symmetry and what is the signal of this scenario are two main subjects worthy of deep exploring.  
The authors of Refs.~\cite{Cline:1999wi,Patel:2013zla} show that  color breaking may precede the color conserving EWPT in the MSSM or  scalar singlets extension of the SM.  The authors of Ref.~\cite{Fornal:2021ovz} show that color breaking vacuum may exist in the Mini-split SUSY.  Furthermore, the authors of Ref.~\cite{Ramsey-Musolf:2017tgh} point out the observed BAU can be realized during the first order color restoration phase transition. Here, we point out that  color breaking may succeed the EWPT, which can be realized in the scenario of electroweak symmetry non-restoration (EWSNR).  In EWSNR models, there are extra singlets coupled to the SM Higgs doublet with negative quartic couplings, which contribute negative thermal masses to the Higgs. If one of them is color triplet, namely, $\Delta$, then the same negative quartic coupling, which gives rise to to a negative squared mass of $\Delta$ in electroweak broken phase, may lead to the spontaneous breaking of the $SU(3)_C$. It is not easy to realize this scenario during the standard EWPT because the VEV of the SM Higgs at finite temperature is smaller than $246$ GeV.  The SM Higgs  may have a much higher VEV  in EWSNR  at  high temperature, which may lower the squared mass  of $\Delta$ to a negative value and thus result in spontaneous symmetry breaking of the $SU(3)_C$. As temperature drops low, the VEV of the SM Higgs  gradually evolve to $246$ GeV and the color symmetry can be restored  by the positive mass term $\mu_c^2 \Delta^\dagger \Delta $ in the potential. We find that both the color symmetry breaking and restoration  processes can be first order, with stochastic GWs as the smoking-gun  of  this two scenarios. Possible signals may be detected by the future GW experiments  such as DECIGO \cite{Kudoh:2005as,Musha:2017usi} and BBO \cite{Cutler:2005qq} for typical benchmarks.

The remaining of the paper is organized as follows: In Section~II, we illustrate the framework.  Section~III is devoted to the calculation of phase transitions.   In Section~IV, we study the stochastic GWs induced by the  first order color breaking and restoration.  The last part is concluding remarks.

\section{The Framework}
We start by presenting the theoretical framework of the color symmetry breaking and restoration.  A key point for the EWSNR is an extra negative thermal mass term for the SM Higgs in addition to the one induced by the plasma of SM particles,
{\small
\bea
\delta V_h(T) \sim {1\over 2 }  c_h T^2 h^2 = {1\over 2}\left[ {3 g^2  +g^{\prime 2}\over 16} + {\lambda \over 2 } + {y_t^2 \over 4}  + \zeta \right] T^2 h^2 , \ \ \ \
\label{thmass}
\eea}
where $g$ and $g^\prime$ are gauge couplings of $SU(2)_L$ and $U(1)_Y$ respectively, $y_t$ is the top quark Yukawa coupling, $\lambda$ the Higgs quartic coupling, $\zeta$ is induced by new Higgs interactions, such as ${1\over 4} \zeta_i s_i^2 h^2 $ or $ {1\over \Lambda } h^2 \bar \chi \chi $ with $s_i$ and $\chi$ are new real scalars and Dirac fermion respectively.  Explicitly, one has 
\bea
\zeta \sim  -{ n_\chi \over 3 } { m_\chi \over \Lambda} - {  n_i \zeta_i \over 12 }
\eea
where $m_\chi$ is the mass of new fermion $\chi$, $\zeta_i$ is the quartic coupling of new scalars, $ n_\chi$ and $n_i $ are degrees of freedom of new species.   At high temperature, positive Higgs thermal mass drive the Higgs VEV to zero, while negative thermal mass pull the Higgs VEV far from zero.  Both $n_\chi$ and $n_i$ are temperature relevant and depend on the UV completion of the theory, which may lead to the electroweak symmetry restoration at super-high temperature.

In this work, we assume that there is a complex color triplet scalar $\Delta$ in the framework EWSNR, with the following effective potential
\bea
V_{\rm eff}^{} =\mu_c^2 \Delta^\dagger \Delta  - \lambda_{hc} H^\dagger H \Delta^\dagger \Delta  +\lambda_c (\Delta^\dagger \Delta)^2   + V_{\rm SNR}^{} 
\eea   
where  $V_{\rm SNR}$ is the effective potential for the  EWSNR, whose expression depends on the model setup and can be found in Refs.~\cite{Meade:2018saz,Baldes:2018nel,Glioti:2018roy,Matsedonskyi:2020mlz,Matsedonskyi:2020kuy,Carena:2021onl,Biekotter:2021ysx,Bai:2021hfb,Matsedonskyi:2021hti}.  Negative quartic coupling $-\lambda_{hc}$ contributes to a negative thermal mass of the SM Higgs as required by the EWSNR, and the same coupling contributes a negative squared mass to the color triplet after electroweak symmetry breaking, which may lead to the spontaneous breaking and restoration of the color symmetry through interfering with the $\mu_c^2$.  

Taking $\sigma$ and $h$ as the background fields of $\Delta$ and $H$, the effective potential at finite temperature can be written as
\bea
\label{background}
V(h,\sigma,T)\supset&& -\frac{\mu_{h T}^{2}}{2}h^2 +\frac{\mu_{c T}^{2}}{2} \sigma^2-\frac{9+2\sqrt{3}}{72 \pi} T g_s^3 \sigma^3 
\\ \nonumber 
&&+\frac{\lambda_{h}}{4}h^4+\frac{\lambda_{c}}{4}\sigma^4-\frac{\lambda_{hc}}{4}h^2\sigma^2, 
\eea
where $g_s$ is the strong coupling, $\mu_{hT}^2=\mu_{h}^{2}-c_h T^2$ and $\mu_{cT}^2=\mu_{c}^{2}+c_c T^2$ with 
\bea
c_c=&&\frac{1}{3}\left(g_s^2+2\lambda_c \right)-\frac{\lambda_{hc}}{6}.
\eea

\begin{figure*}[tbp]
	\includegraphics[width=9cm]{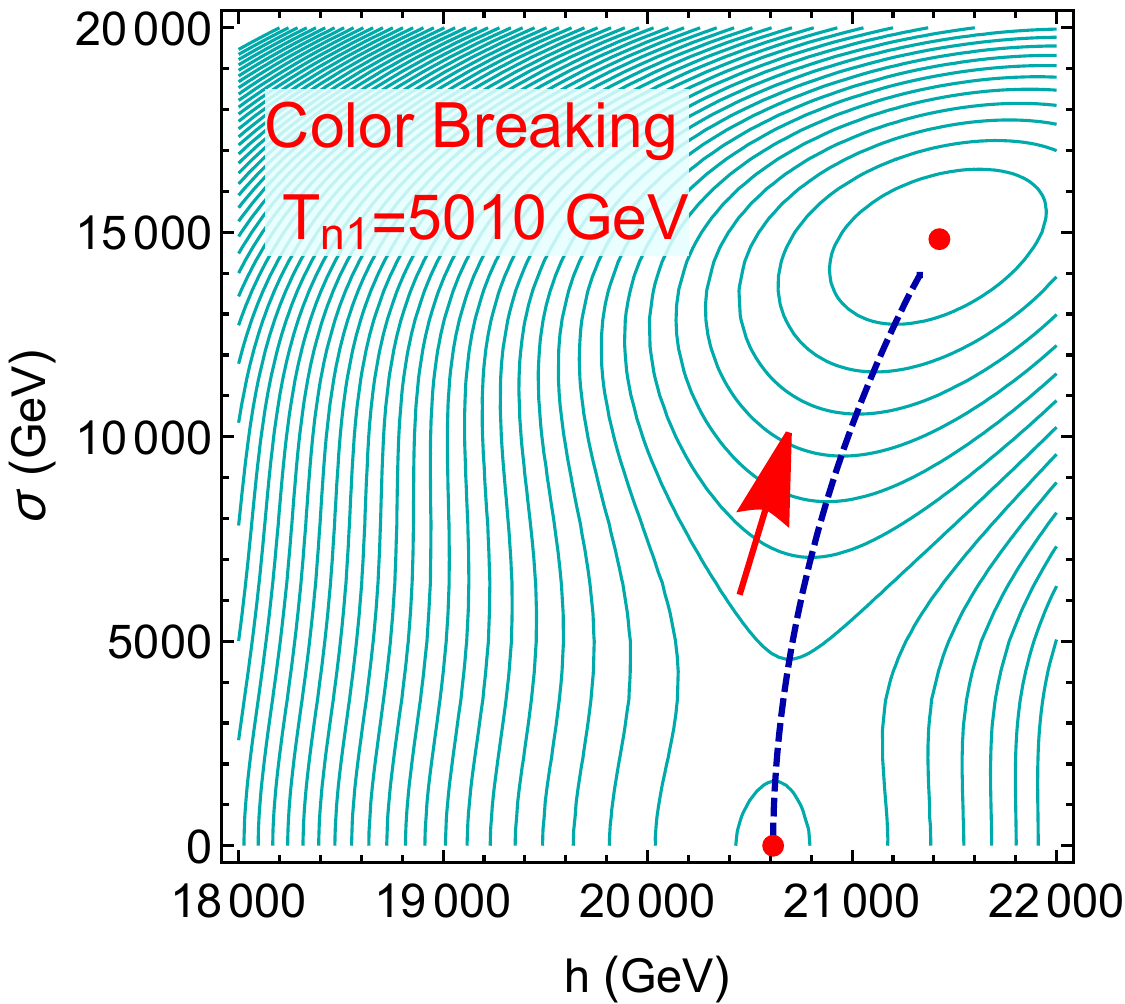}
	\includegraphics[width=8.5cm]{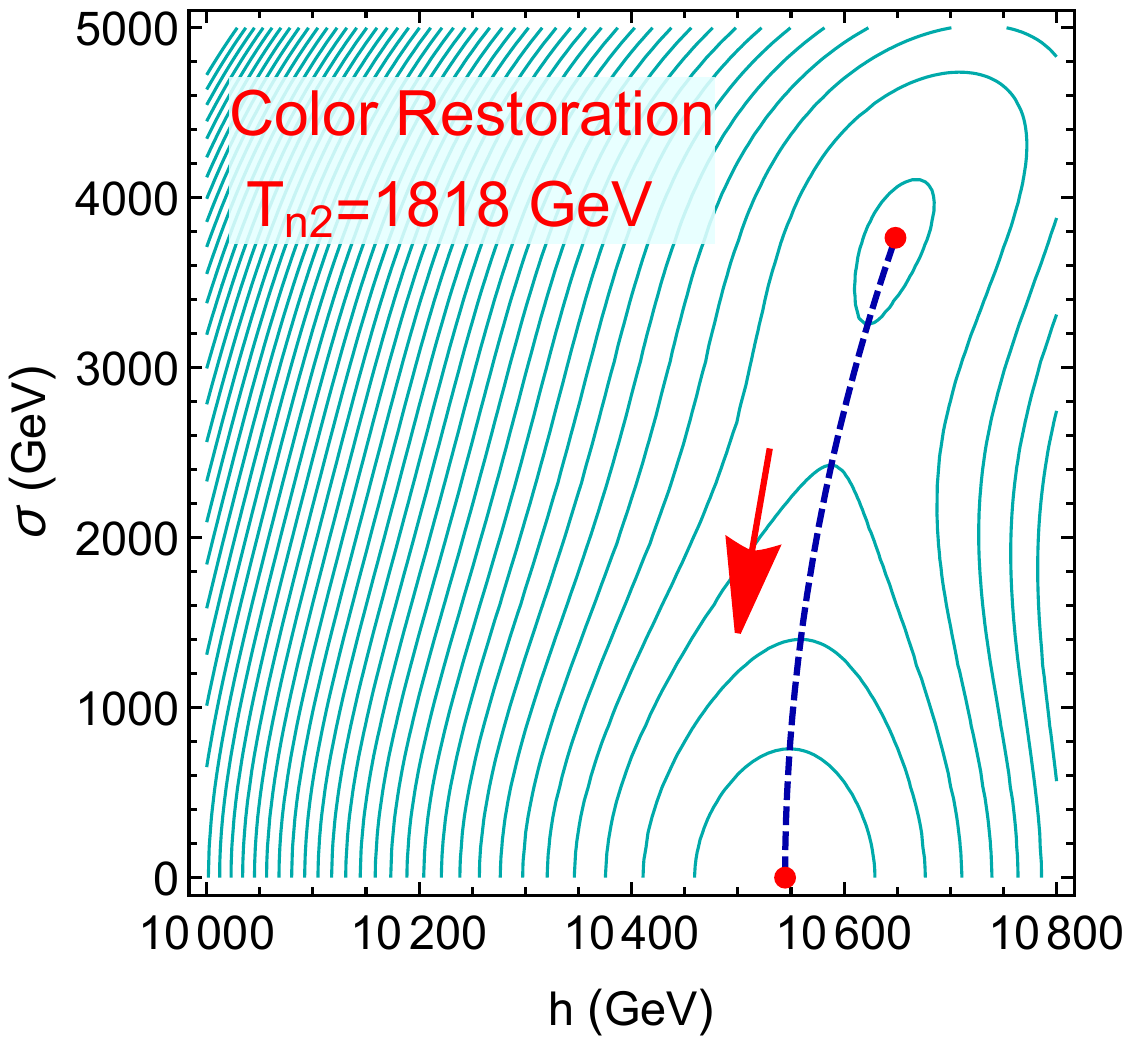}
	\caption{(color online). Contours of the free energy in solid cyan lines and trajectory of the bounce field in dashed darker blue line for the benchmark point shown in TABLE \ref{table1}. Red points stand for the two minima.}
	\label{contour2}
\end{figure*}

\section{Color symmetry breaking and restoration}

In this section, we check the pattern of color symmetry breaking and restoration processes. As can be seen from the Eq.~(\ref{background}), the cubic term, which is induced by the gluon fields,  may lead to a first order phase transition, which is completed by the nucleation, expansion and coalescence of the bubbles, with the bubble nucleation rate  per unit time per unit volume given by~\cite{Turner:1992tz}
\begin{eqnarray}
\Gamma(T)\simeq T^4 \left( \frac{S_3}{2\pi T} \right)^{3/2}e^{-S_3/T},
\end{eqnarray}
where $S_3$ is the three-dimensional Euclidean action taking the following form
\begin{eqnarray}
S_3 = \int_{0}^{\infty} dr\ r^2 \left[\frac{1}{2}\left(\frac{d\vec{\phi}(r)}{dr}\right)^2 + V(\vec{\phi},T)\right] \; .
\end{eqnarray}
where $\vec{\phi}=(h,\sigma)$. The corresponding equation of motion is
\begin{eqnarray}
\frac{d^2\vec{\phi}}{dr^2}+\frac{2}{r}\frac{d\vec{\phi}}{dr}=\frac{dV(\vec{\phi},T)}{dr} ,
\end{eqnarray}
with the bounce boundary conditions
\begin{eqnarray}
\lim\limits_{r\rightarrow \infty} \vec{\phi}(r)=0,\quad \frac{d\vec{\phi}}{dr}\Bigr|_{r=0}=0.
\end{eqnarray}
The condition for bubble nucleation is that one bubble is formed per unit horizon volume at the temperature $T_n$, which gives~\cite{Breitbach:2018ddu,Cai:2017tmh,Eichhorn:2020upj}
\begin{eqnarray}
\frac{S_3}{T}
&&\approx ~ 146-2~ {\rm log}\left(\frac{g_*}{100}\right)-4~ {\rm log}\left(\frac{T_{n}}{100\rm GeV}\right) \; , 
\label{S3}
\end{eqnarray} 
where $g_*$ is the relativistic effective degrees of freedom at $T$. For nucleation temperatures  $T_{n1}\sim 5\times 10^3 ~{\rm GeV}$ and $T_{n2}\sim 2\times 10^3 ~{\rm GeV}$, one has  $S_3/T\sim 130.2$ and $S_3/T\sim 133.9$, respectively.

To study the pattern of color breaking/restoration, we set 
\bea
\zeta = -\xi {4\over 2 + T/{\rm TeV}  }
\eea
where parameter $\xi=4.5$. This scenario can be realized in a model  of SNR assisted  by scalars with different masses, where the contribution of a new scalar to the Higgs thermal mass loses efficacy when temperature drops below its mass. Parameters for the benchmark point are listed in TABLE~\ref{table1}.

The left-panel of Fig.~\ref{vev-hc} shows the schematic diagram for  the phase transitions. First, the SM Higgs and the color triplet are located at  the symmetric phase $(0,0)$ in the early Universe. With the decrease of temperature, the electroweak symmetry is first broken. After that, the color symmetry is broken at a lower temperature and both two fields have non-zero VEVs $(h,\sigma)$.  Finally, with the further decrease of the temperature, the color symmetry is restored  and only  the SM Higgs has a non-zero VEV $(h,0)$.  The right-panel of the Fig.~\ref{vev-hc} shows the VEVs of the SM Higgs and the color triplet as the function of temperature for the benchmark point in the TABLE~\ref{table1}. It is clear that the EWPT takes place at $T_0=44017~\rm GeV$. After that, the first order color symmetry breaking occurs at $T_{n1}=5010 ~\rm GeV$ (see the dashed vertical brown line), at which both VEVs are discontinuous. The cyan shadow represents the region where the color symmetry is broken.  Finally,  the color symmetry is restored at $T_{n2}=1818~\rm GeV$ and the corresponding process is also first order (see the dashed vertical green line). When calculating the bubble nucleation temperatures we use the public package \texttt{FindBounce}\cite{Guada:2020xnz}.

\begin{figure*}[t!]
	\includegraphics[width=8.6cm]{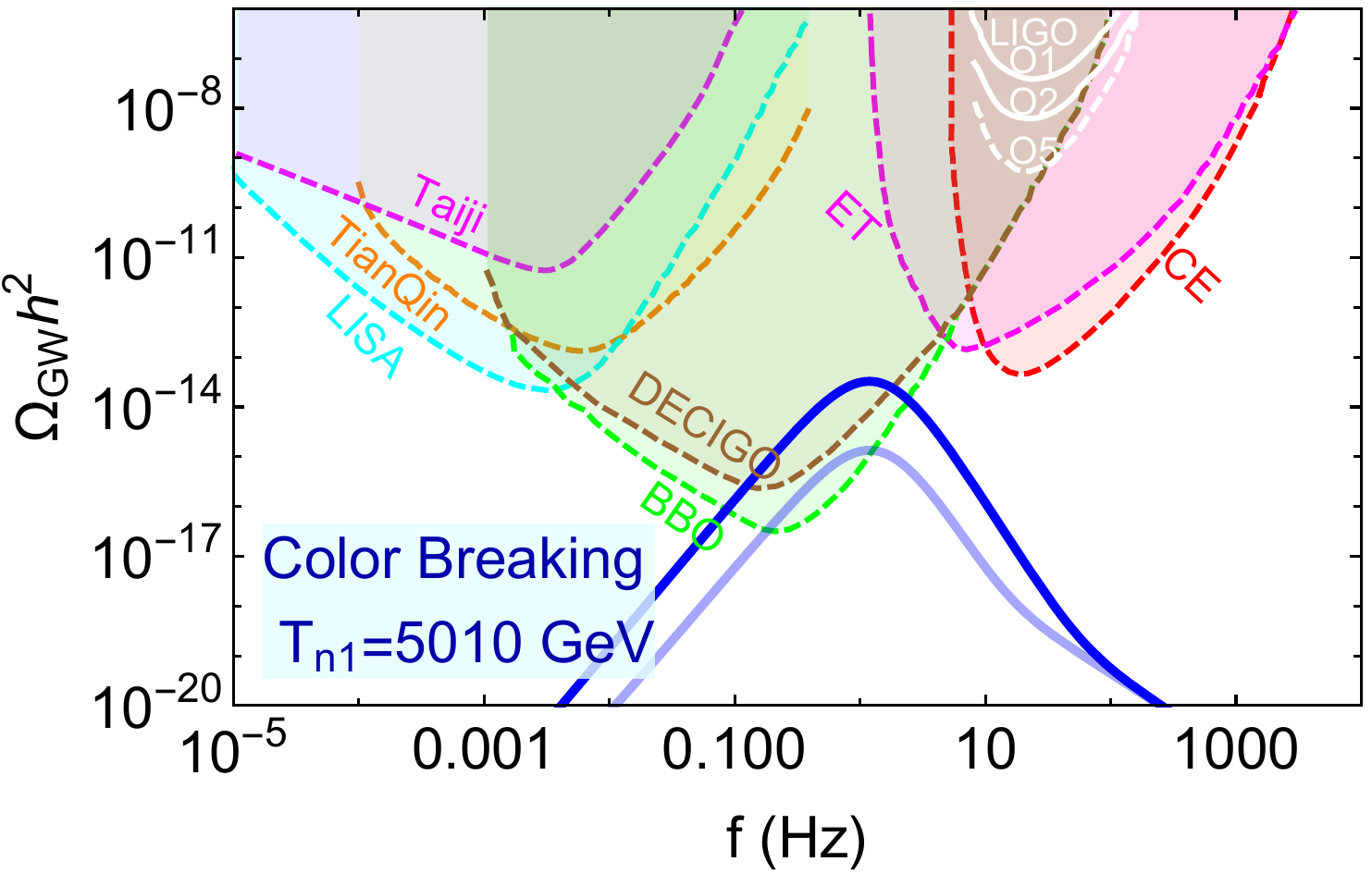}
	\includegraphics[width=8.6cm]{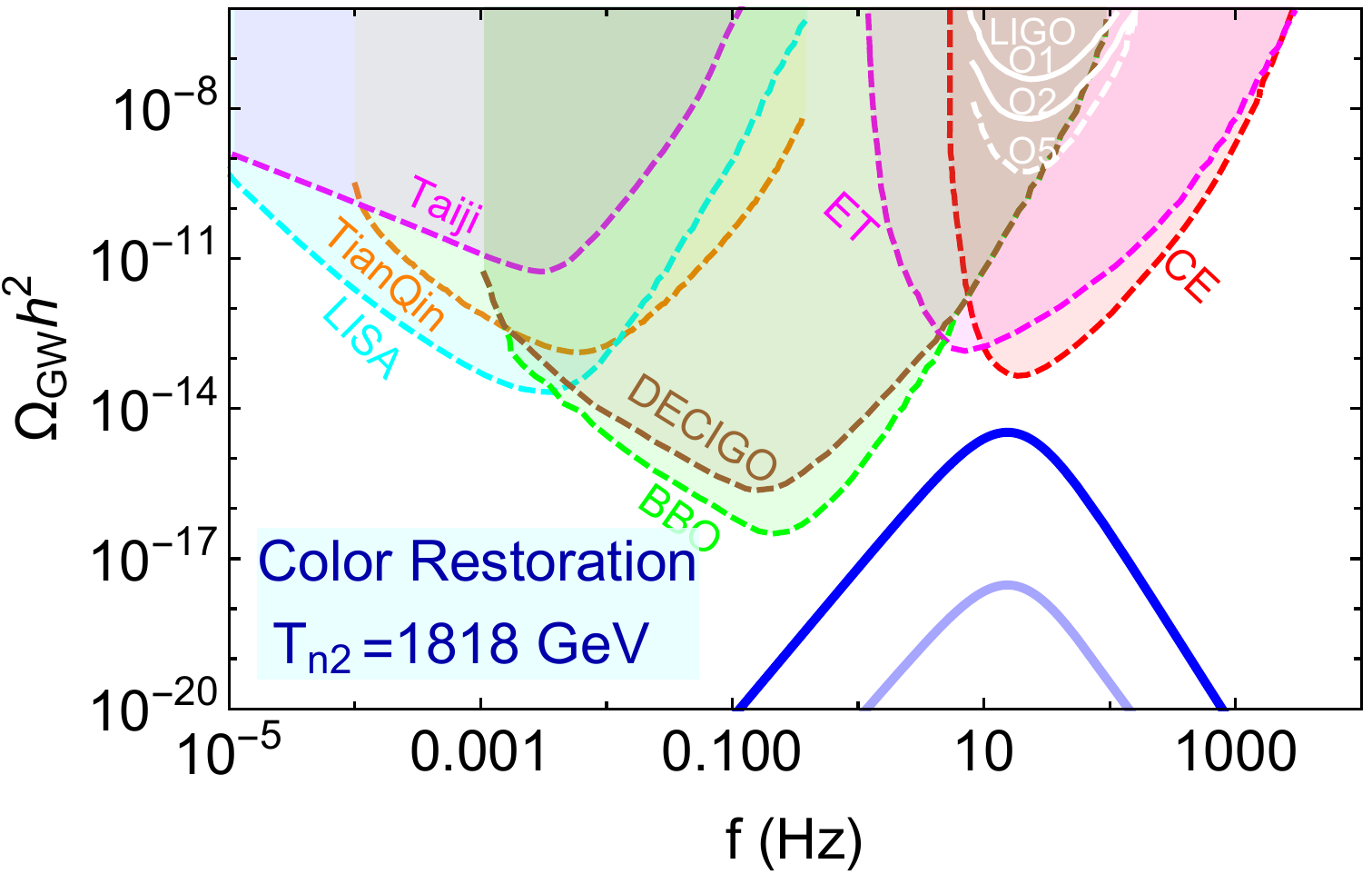}
	\caption{
     GW  spectrum without the factor $\Upsilon$ (dark blue lines) and with $\Upsilon$ (light blue lines) for the first order color breaking (left-panel) and restoration (right-panel)  phase transitions with $v_w=0.63$. The benchmark point is given in TABLE~\ref{table1}.}
	\label{GW2}
\end{figure*}

Fig.~\ref{contour2} shows the contours of the free energy in the $(h, ~\sigma)$ plane at the color symmetry breaking (left-panel) and restoration (right-panel) temperatures. The solid cyan lines represent contours of the free energy and the blue dashed lines represent the trajectory of the bounce field. 

\begin{table}[!htp]
	\footnotesize
	\begin{tabular}{c c c c c c c c}
		\hline
		$\mu_c$(GeV)~ & ~ $\lambda_c$ ~ & ~ $\lambda_{hc}$ ~ & ~ $\xi$ ~ & ~ $T_{n1}(T_{n2})$~ & ~ $\alpha$~ & ~ $\beta/H_{n}$  \\
		\hline
		$1225.8$ & $0.051$ & $0.04$ & $4.5$ & $5010 (1818)$ & $0.02 (0.03)$ & $776 (27723)$ \\
		\hline
	\end{tabular}
	\caption{Parameters for the benchmark point and phase transition parameters for the color symmetry breaking (restoration) processes.}
	\label{table1}
\end{table}

\section{Stochastic gravitational waves}
There are generally three sources for GW production during a cosmological first order phase transition: (a) from the stress
energy concentrated at the bubble wall, called the bubble collisions~\cite{Huber:2008hg,Di:2020nny}; (b) from the acoustic production due to
the kinetic energy of the plasma in the form of sound waves~\cite{Hindmarsh:2015qta}; (c) from the turbulent motion of the plasma called the 
magenetohydrodynamic (MHD)~\cite{Binetruy:2012ze,Caprini:2009yp}. For a phase transition in a thermal plasma, as is the case here, the contribution
from sound waves dominates with MHD comes next while the one from bubble collisions is negligible. We thus consider only the contributions from sound
waves and MHD, in which case the GW spectrum today can be approximated as a linear summation of them:
\begin{equation}
\Omega_{\rm GW}(f)h^2\approx\Omega_{\rm sw}(f)h^2+\Omega_{\rm turb}(f)h^2.
\end{equation}

For all cases, the resulting spectra of GWs depend on the
parameters characterizing the bulk properties of the system. 
One is the energy released during the transition, which when normalized by the total radiation energy density at $T_n$ corresponds to a dimensionless
parameter:
\begin{eqnarray}
\alpha=\frac{1}{g_*\pi^2T_n^4/30}\left(\Delta V_T-\frac{T}{4}\frac{\partial\Delta V_T}{\partial T}\right)\Big|_{T_n}.
\end{eqnarray}
Another is a inverse time scale $\beta$ and similarly a dimensionless ratio can be defined:
\begin{eqnarray}
\frac{\beta}{H_n}=T_n\frac{d(S_{3}/T)}{dT}\Big|_{T_n},
\end{eqnarray}where the Hubble rate $H$ is evaluated at a temperature close to $T_n$. This quantity is typically used as it is
related to the characteristic length scale in the transition, the mean bubble separation $R_{\ast}$ which determines the peak of the GWs.
For an exponential nucleation of bubbles, when the quantity $S_3/T$ has no minimum near $T_n$, these two parameters are related~\cite{Hindmarsh:2019phv},
\begin{eqnarray}
H R_{\ast} = (8 \pi)^{1/3} \frac{v_w}{\beta/H}.
\end{eqnarray}In principle $\beta$ depends on the bubble wall velocity $v_w$. Here for simplicity we will use the definition above.
The wall velocity $v_w$ is another crucial parameter. However its calculation is still a work going on. Here we simply choose it to be a free parameter.

There have been significant progress on numerical simulations carried out for acoustic production, with a spectrum given by~\cite{Hindmarsh:2015qta}
\begin{eqnarray}
\label{sw}
&& \Omega_{\rm sw}(f)h^2=2.65\times10^{-6}\frac{1}{\beta/H_n}\left(\frac{\kappa_v\alpha}{1+\alpha}\right)^2\left(\frac{g_*}{100}\right)^{-1/3} \nonumber \\
&& \quad \quad \quad \times v_w\left(\frac{f}{f_{\rm sw}}\right)^3\left(\frac{7}{4+3(f/f_{\rm sw})^2}\right)^{7/2} \Upsilon(\tau_{\text{sw}}), \ \  
\end{eqnarray}
where $f_{\text{sw}}$ is the peak frequency seen today
\begin{equation}
\label{swf}
f_{\rm sw}=1.9\times10^{-5}~{\rm Hz}\times\frac{\beta/H_n}{v_w}\left(\frac{T_n}{100~{\rm GeV}}\right)\left(\frac{g_*}{100}\right)^{1/6}.
\end{equation}
and $\kappa_{v}$, typically a function of $\alpha$ and $v_w$, is the fraction of the energy released that goes to the kinetic energy of the sound waves. 
The functional dependence $\kappa_v(\alpha, v_w)$ can be obtained by numerically solving a set of hydrodynamic equations governing the evolution of the plasma, 
which, in the form of a perfect fluid, admits three modes, deflagration, supersonic deflagration (also called hybrid), and detonation when $v_w$ increases~\cite{Espinosa:2010hh}.
The result can be fitted well by a set of analytical formulas~\cite{Espinosa:2010hh} and we use it here. Recent numerical simulations, however, show that this result 
from the above-mentioned hydrodynamic modelling cannot describe well the entire parameter space of $v_w$ and $\alpha$. For small $v_w$ and large $\alpha$, 
there is a deficit of GWs produced from simulations, attributable to the reduced $\kappa_v$ compared with the value from the analytical modelling~\cite{Cutting:2019zws}.

In addition, the factor $\Upsilon$~\cite{Guo:2020grp}, which is a function of the lifetime of the sound waves $\tau_{\text{sw}}$, corrects for the finite lifetime~\cite{Ellis:2020awk} 
of the sound waves in an expanding universe which was taken to be infinite in previous studies. Physically it describes the increasingly attenuated 
contribution as the source is being diluted as the universe expands. Its value thus depends on how the universe expands when the sound waves is
active. For a radiation dominated universe, as is the case here, it is given by~\cite{Guo:2020grp}
\begin{eqnarray} 
\Upsilon = 1 - \frac{1}{\sqrt{1 + 2 \tau_{\text{sw}} H_n}},
\end{eqnarray}The problem thus remains on the calculation of the lifetime of the sound waves $\tau_{\text{sw}}$. It is usually approximated to be 
the time scale for the onset of turbulence, given by~\cite{Hindmarsh:2015qta}
\begin{eqnarray}
\tau_{\text{sw}} \approx \frac{R_{\ast}}{\bar{U}_f},
\label{eq:tausw}
\end{eqnarray}where $\bar{U}_f=3 \kappa_v\alpha/(4(1+\alpha))$~\cite{Weir:2017wfa} which is the mean fluid velocity.

The GW spectrum from the MHD is probably the least understood compared with the other two sources,
and recent results from numerical simulations strongly depend on the initial conditions~\cite{Caprini:2015zlo,Kahniashvili:2008pf,Kahniashvili:2008pe,Kahniashvili:2009mf,Caprini:2009yp,Kisslinger:2015hua,Pol:2019yex}.
We adopt the spectrum estimated from an analytical understanding~\cite{Binetruy:2012ze,Caprini:2009yp} and caution
about its potential significant changes in the future. This spectrum is
\begin{eqnarray}
\label{turb}
\Omega_{\rm turb}(f)h^2=3.35\times10^{-4}\frac{v_w}{\beta/H_n}\left(\frac{\kappa_{\rm turb}\alpha}{1+\alpha}\right)^{3/2}\\  \nonumber
\times\left(\frac{g_*}{100}\right)^{-1/3}\frac{(f/f_{\rm turb})^3}{\left[1+(f/f_{\rm turb})\right]^{11/3}(1+8\pi f/h_*)}, 
\end{eqnarray}
where the efficiency factor $\kappa_{\rm turb}$ is unknown and we take $\kappa_{\rm turb} \approx (5\sim10)\%~\kappa_{v}$ ~\cite{Hindmarsh:2015qta}. The peak frequency $f_{\rm turb}$ is

{\small
\bea
f_{\rm turb}=2.7\times10^{-5}{\rm Hz}\times\frac{\beta/H_n}{v_w}\left(\frac{T_n}{100{\rm GeV}}\right)\left(\frac{g_*}{100}\right)^{1/6},
\eea}

and 
\begin{eqnarray}
h_*=16.5\times10^{-6}~{\rm Hz}\left(\frac{T_n}{100~{\rm GeV}}\right)\left(\frac{g_*}{100}\right)^{1/6}.
\end{eqnarray}

Fig.~\ref{GW2} shows the sensitivities of GW detectors and GW spectrum for the benchmark point in TABLE \ref{table1} and choosing the following bubble wall velocity $v_w=0.63$. The dark and light blue solid lines represent the GW  spectrum without the factor $\Upsilon$ and with the factor $\Upsilon$, respectively. The  dashed lines are the power-law integrated sensitivity curves for the Taiji \cite{Guo:2018npi}, LISA \cite{Audley:2017drz,Breitbach:2018ddu}, TianQin \cite{Mei:2020lrl}, BBO \cite{Cutler:2005qq}, DECIGO \cite{Kudoh:2005as,Musha:2017usi}, ET \cite{Hild:2010id,Punturo:2010zz}, CE \cite{LIGOScientific:2016wof} and LIGO \cite{LIGOScientific:2014qfs,LIGOScientific:2019vic}.
We can see for this benchmark the GW signal from the color breaking transition is within reach of DECIGO and BBO when $\Upsilon\approx 1$ though it is weaker
when using Eq.~(\ref{eq:tausw}). However the value of $\tau_{\text{sw}}$ and thus $\Upsilon$ is still highly uncertain and its determination needs
considerable numerical simulations and analytical insights in the future. In addition, a full exploration of the parameter space will potentially
lead to more promising regions for detectable  stochastic GW signal.

\hspace{1cm}

\section{Summary and outlook}\label{5}
Symmetries are essential for conservation laws. It is well-known that the $SU(3)_C$ and $U(1)_{\rm em}$  are good symmetries at zero-temperature. However, they can be broken at high temperature as first pointed out by S.Weinberg. In this letter, we investigated the color symmetry breaking and restoration in the EWSNR framework. Our study shows that both processes can be first oder, with stochastic GWs as the smoking-gun, which can be detected in further GW experiments such as DECIGO and BBO.  Our study provides a new approach for color breaking Baryogenesis  at the TeV scale  as well as  new sources of stochastic GWs  with two different peak frequencies, which is the typical signal of this scenario.

\section*{Acknowledgments}
This work was supported by the National Natural Science Foundation of China under grant No. 11775025 and No. 12175027.

\bibliography{CoB-SNR}
\end{document}